# The average free volume model for liquids

Yang Yu*[a,] Reinhard Krause-Rehberg[b]

[a]School of Physics and Optoelectronic Engineering, Nanjing University of Information Science & Technology, Nanjing 210044, Jiangsu, China

[b]Institut für Physik, Martin-Luther-Universität Halle-Wittenberg, Von-Danckelmann-Platz 3, 06120 Halle, Germany

*Author to whom correspondence should be addressed. Electronic mail: yuyang5020@googlemail.com

**Abstract:**

In this work, the molar volume thermal expansion coefficient of 59 room temperature ionic liquids is compared with their van der Waals volume $V_w$. Regular correlation can be discerned between the two quantities. An average free volume model, that considers the particles as hard core with attractive force, is proposed to explain the correlation in this study. A combination between free volume and Lennard-Jones potential is applied to explain the physical phenomena of liquids. Some typical simple liquids (inorganic, organic, metallic and salt) are introduced to verify this hypothesis. Good agreement from the theory prediction and experimental data can be obtained.

Keywords: Average free volume; ionic liquids; Lennard-Jones potential; simple liquids.

**Introduction:**

Comparing to the gas and the crystalline solid, the nature of the liquid is still unclear. There is no theory for the universally prediction of the properties of liquids from microscopic structure. On the other side, the liquid materials share many common phenomena. Normally $T_g/T_m$ ($T_g$ is glass transition temperature, $T_m$ is melting temperature) is about 2/3, which fits the Boyer-Beaman rule for polymers, ionic liquids and even inorganic glass[1-4]. The Eötvos and Guggenheim empirical equations establish the relationship between surface tension, liquid density and critical temperature for the majority of liquids[5,6]. For the polymers, there is a correlation between crystalline volume $V_c$ and van der Waals volume $V_w$: $V_c \sim 1.435 V_w$[7]. For the ionic liquids, it is $V_c$ (=$V_M$) ~ 1.410 $V_w$ ($V_M$ is molecular volume)[8]. The values 1.435 for polymers and 1.410 for ionic liquids are close to $2^{1/2}$, which corresponds to the minimum energy in Lennard-Jones (or 6-12) potential (at the position $r/\sigma=2^{1/6}$, then the volume ratio is $(2^{1/6})^3=1.414$)[8-10]. The common works of liquids focus on the dynamics of particles and harsh repulsion within short range. And normally the attractive interaction is considered as introducing uniform background potential that provides the cohesive energy[11]. There are few studys working on the free space between molecules[12-16]. In this study, according to the relationship between thermal expansion coefficient and van der Waals volume for up to 59 ionic liquids, a new average free volume model – that is considering the atom or molecule as hard core with attractive force, each particle is surrounded by the average free volume – is established[16]. To prove the validity of this new model, some typical simple liquids are introduced for discussion and analysis.

**Results and discussion:**

From the experimental data[1,17], under the atmospheric pressure, the molar volume $V_{mol}$ displays linear correlation with temperature $T$ for all ionic liquids, which can be presented by the linear function:

$$V_{mol}(T) = V_{E0} + C_1 T \qquad (1)$$

Here, $V_{E0}$ is the volume when extrapolate the molar volume at liquid state to absolute zero. The constant $C_1$ corresponds to the thermal expansion coefficient of the molar volume. Up to 59 ionic liquids are fit to equation (1), the experimental data are taken from the NIST website [17,18]. The fitting results of $V_{E0}$ and $C_1$ are compared with the van der Waals volume $V_w$ of the ionic liquids[8,19,20]. The van der Waals volume is the space occupied by a molecule, which is impenetrable to other molecules with normal thermal energies[10,21,22]. For more details, see table 1 in SI. Obvious correlation between $V_{E0}$, $C_1$ and $V_w$ can be discerned from figure 1.

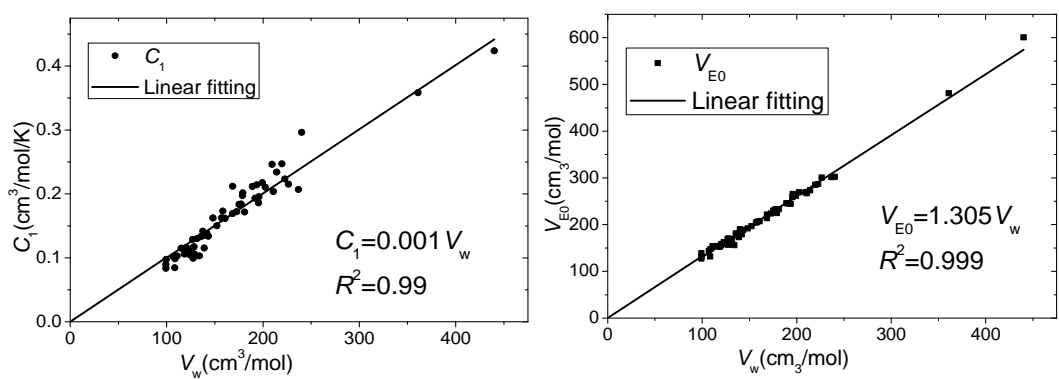

**Fig. 1** Correlation between $C_1$, $V_{E0}$ and $V_w$. The data are fitted by linear function fixing intercept at 0. $C_1$: thermal expansion coefficient of molar volume. $V_{E0}$: extrapolation of molar volume from liquid state to absolute zero. $V_w$: van der Waals volume.

From the fitting result, $C_1$ shows a correlation with $V_w$. $V_{E0}/V_w=1.305V_w$. This value is smaller than crystalline volume ratio $V_c/V_w= 1.410$ for the ionic liquids[8], and it is close to the value of dense packing for crystal 1.35 (reciprocal of 74.05%. When cancel the intercept limitation in the fitting of $V_{E0}$ to $V_w$, the slope is 1.34).

**Proposed model:**

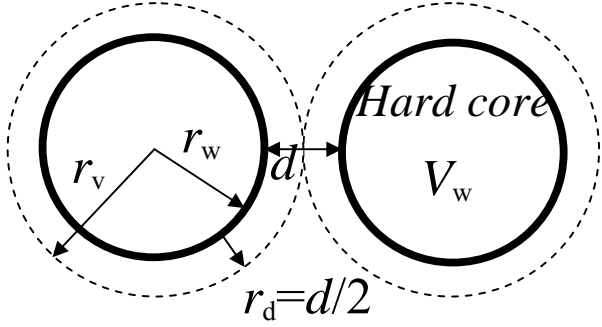

**Fig. 2** Schematic average free volume model for the liquids. $r_w$ is van der Waals radius. $r_V$ is the radius of the larger sphere. $d$ is distance between the two particles. $r_d$ is half of $d$.

The correlation between $C_1$, $V_{E0}$ and $V_w$ can be explained by a schematic average free volume model displayed in figure 2. The molecules are simplified as spheres. Two particles are surrounded by the average free volume, and separated by distance $d$, then the occupied volume for each particle $V$ is supposed to be the larger sphere volume (hard core volume $V_w$ adds the average free volume) plus interstitial volume $V_I$: $4\pi/3 * r_V^3 + V_I$, here $r_V = r_w + r_d$. So $V = 4\pi/3*(r_w^3 + 3r_w^2 r_d + 3r_w r_d^2 + r_d^3) + V_I$. When temperature increases, $r_d$ increases, so $r_d$ is a function of $T$, $r_d(T)$. For simplicity, the first order approximation is applied to the formula, second and third order of $r_d$ are omitted[16]. Then:

$$V(T) = V_w + V_I + 4\pi r_w^2 r_d(T) \qquad (2)$$

The comparison between equation (1) and (2) indicates the relationship:

$$V_{E0} = V_w + V_I \qquad (3)$$

Because the molar volume $V_{mol}$ linearly increases with temperature, so $r_d$ linearly changes with $T$, that is $r_d = C_2 T$, $C_2$ is constant:

$$C_1 = 4\pi r_w^2 C_2 = 4\pi C_2 [3V_w/(4\pi)]^{2/3} \qquad (4)$$

The $V_{E0}$ equals the van der Waals volume plus the interstitial volume, corresponding to the dense packing of crystal. And the thermal expansion coefficient $C_1$ has a positive correlation with van der Waals volume $V_w$.

Here, the average free volume $v_f$ is the free space averaged to each molecule, it is different from the local free (hole) volume $v_h$, which is the cavity in the real structure that can be seen from the positron annihilation lifetime spectroscopy (PALS) experiment. The hole volume $v_h$ is generated from coalescence of $v_f$ with statistics possibility because the dynamic movement of particles[23-25].

The $C_2$ depends on the force and molecular movement between particles. Here the sphere is considered for simplification, when real particles are introduced, molecular structure and configuration should be accounted in. These are the reasons for the dispersion in the fitting of figure 1.

According to the above model, the volume of liquids can be represented as $V=C_1T+1.35V_w$. At the melting point $T_m$, $V=V_{melt}$, then gives the formula: $V=(V_{melt}-1.35V_w)T/T_m+1.35V_w$. Then at the boiling point, $T_b=T_m*(V_b/V_w -1.35)/(V_{melt}/V_w -1.35)$.

To verify this conclusion, experimental data for 24 different typical simple liquids are applied[26]. Detailed information can be found in the SI table 2.

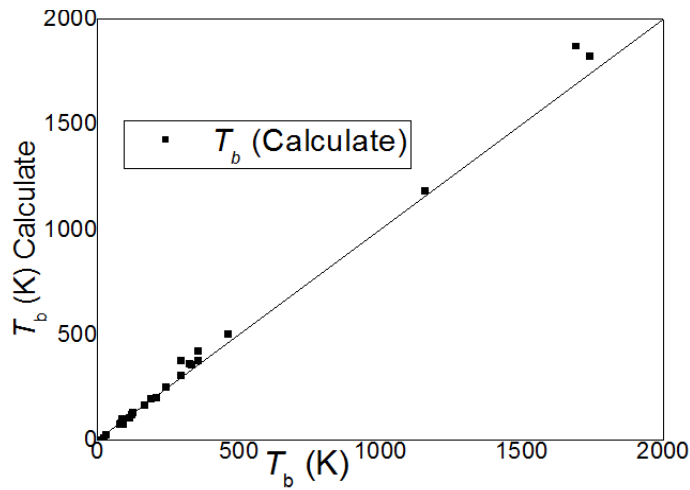

**Fig. 3** Comparison of boiling temperature between experiment and calculation from the average free volume model. The line is a guide to the eye. Simple liquids introduced here are: He_4, Ne, Ar, Kr, Xe, CO, $H_2$, NO, $N_2$, $O_2$, HCl, $Br_2$, $Cl_2$, $F_2$, HBr, $I_2$, $CH_4$, $NO_2$, Na, NaCl, KCl, $C_6H_6$, $C_6H_{12}$ and $C_5H_{10}$.

Two main factors are not considered in this model: magnitude of the attractive force and the molecule shape. Intuitively, particles with larger force make it harder to be separated, then $C_2$ is smaller. To verify this assumption, the potential energy change with temperature in one degree is calculated and some typical simple liquids are introduced for analysis.

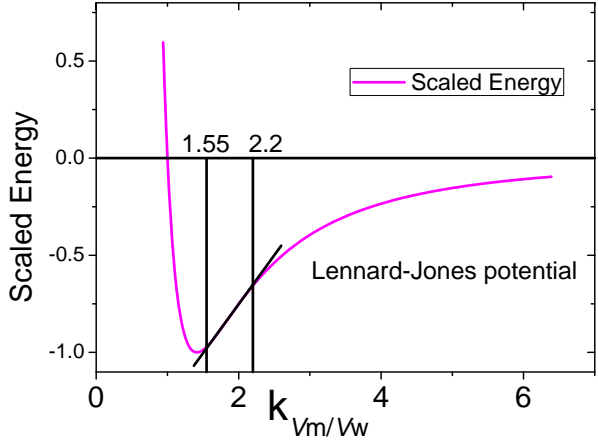

**Fig. 4** Lennard-Jones potential as a function of volume ratio $k_{V\text{mol}/V\text{w}}$. For the liquids, the volume ratio between $V_{\text{mol}}$ and $V_{\text{w}}$ normally falls between the range 1.55 to 2.2. In this range, the scaled energy near-linearly changes with volume ratio.

The interaction between a pair of neutral atoms or molecules can be approximates by the Lennard-Jones potential formula[9]:

$$V(r) = 4\varepsilon[(\frac{\sigma}{r})^{12} - (\frac{\sigma}{r})^{6}] \qquad (5)$$

Here $\varepsilon$ is the depth of the potential well, $\sigma$ is the collision diameter at which the inter-particle potential is zero, $r$ is the distance between the particles. In this work, $\sigma$ can be displaced by van der Waals radius $r_{\text{w}}$, then $(\sigma/r)^3 = (2r_{\text{w}}/r)^3 = V_{\text{w}}/V_{\text{mol}}$. Set $k_{V\text{mol}/V\text{w}} = V_{\text{mol}}/V_{\text{w}}$ Then the Lennard-Jones potential can be written as:

$$\frac{V(r)}{\varepsilon} = 4[(\frac{1}{k_{V_{mol}/V_w}})^{4} - (\frac{1}{k_{V_{mol}/V_w}})^{2}] \quad (6)$$

For the majority liquids, between the melting temperature $T_{\text{m}}$ and boiling temperature $T_{\text{b}}$, the volume ratio of molar volume to van der Waals volume is in the range around 1.6. The minimum is not smaller than 1.45 and maximum not larger than 2.3 (only with exception of helium He and hydrogen $H_2$) [27]. When taking the range 1.55 ~ 2.2, the Lennard-Jones potential energy nearly linearly increase with $k_{Vmol/Vw}$ as shown in figure 4. This indicates that the linear change of volume with temperature roots in the linear change of potential energy with temperature.

The potential energy change can be presented in terms of surface tension as displayed in figure 5. The surface tension $\gamma$ expands the free volume, changes the sphere surface from $S_1$ to $S_2$, then the energy change with temperature in one degree is:

$$\Delta E = \gamma S_2 - \gamma S_1 = \gamma 4\pi (r_w + r_d + \Delta r_d)^2 - \gamma 4\pi (r_w + r_d)^2 \\ = \gamma 4\pi [2\Delta r_d (r_d + r_w) + \Delta r_d^2] \quad (7)$$

Since $\Delta r_d \ll r_w$, $r_d \ll r_w$ the terms $\Delta r_d^2$ and $r_d$ are omitted. $\Delta r_d = C_2 \Delta T$, $\Delta T = 1$ K. So:

$$\Delta E = 8\pi \gamma C_2 r_w \quad (8)$$

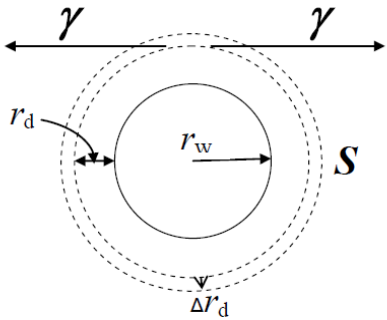

**Fig. 5** Energy change from the free volume expansion by the surface tension.

Combining equation (4) and (8):

$$\Delta E = \frac{2\gamma C_1}{[3/(4\pi)]^{1/3} V_w^{1/3}} \quad (9)$$

The thermal expansion coefficient $C_1$ is reciprocally proportional to surface tension for the liquids. Since the surface tension is measure of the force. The intuition that the thermal expansion coefficient $C_1$ is smaller when interaction is stronger between particles is verified, if $\Delta E$ is the same for all the liquids.

To verify whether $\Delta E$ is the same, some typical simple liquids are chosen for comparison. When calculating the molar volume thermal expansion coefficient $C_1$, only the temperature range near $T_m$ where the molar volume linearly expanded with $T$ is chosen. Normally this it is the range between $T_m$ and $T_b$.

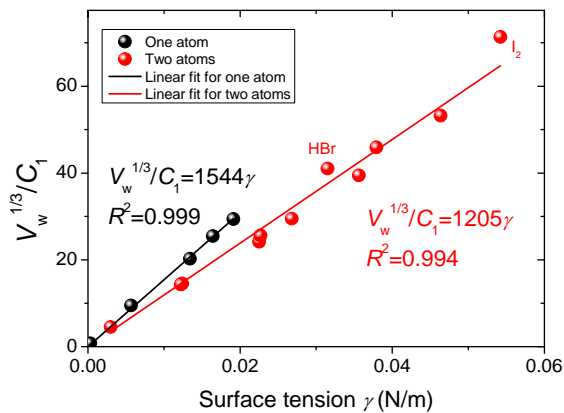

**Fig. 6** Comparison between $V_w^{1/3}/C_1$ and surface tension $\gamma$ for one atom and two atoms liquids. The units for $V_w$, $C_1$ and $\gamma$ are cm$^3$/mol, cm$^3$/mol/K and N/m in this figure. One atom liquids are: He_4, Ne, Ar, Kr and Xe. Two Atoms liquids are: CO, $H_2$, NO, $N_2$, $O_2$, HCl, $Br_2$, $Cl_2$, $F_2$, HBr and $I_2$. Detailed information can be found in the SI, table 2.

As shown in figure 6. For the one atom (noble gas He_4, Ne, Ar, Kr, Xe) line, the slope of the linear fitting is 1544, that gives the energy change under one degree 17.63 J/mol/K. That is $2.12R$. Here $R$ is gas constant. For the two atoms line (CO, $H_2$, NO, $N_2$, $O_2$, HCl, $Br_2$, $Cl_2$, $F_2$, HBr, $I_2$ ), it is 22.59 J/mol/K. That is $2.72R$. For the liquids under 1 atm pressure, the heat capacity under constant pressure $C_p$ for the noble gas is around 40

J/mol/K[17,18], that is about $5R$, and for the other liquids as $O_2$, $N_2$ and etc., this value is about 56 J/mol/K[17,18], that is around $7R$.

Other liquids ($CH_4$, $NO_2$, Na, NaCl, KCl, $C_6H_6$, $C_6H_{12}$, $C_5H_{10}$, $C_8H_{15}BF_4N_2$) are also introduced in this work. As displayed in the SI table 2, the energy change for all the liquids in the study is around $2R \sim 3R$. This value is near the kinetic energy of particles 3/2R.

According to the result and discussion above, kinetic energy gives rise to potential energy change for each particle, by means of repulsive force. The hard core with kinetic energy pushes away other particles, that expands the volume.

**Conclusion:**

According to the correlation between molar volume thermal expansion coefficient, extrapolate volume at absolute zero and van der Waals volume, an average free volume model is proposed. The volume can be estimated only with the information of van der waals volume for ionic liquids: $V_{mol}=0.001V_wT+0.13V_w$. For the simple liquids, when known the volume at melting point and melting temperature, $V_{mol}=(V_{melt}-1.35V_w)T/T_m+1.35V_w$.

The ratio of the molar volume to van der Waals volume for liquids is around the range 1.5 ~2.2. From the Lennard-Jones potential, in this range, the scaled potential can be approximated as linearly change with volume ratio. Then linear function of volume to temperature means linear change of potential energy, this energy change with temperature in one degree is around $2R$ to $3R$ for the simple liquids in this study.

This average free volume model is more a technique method than corresponding to a real structure, since from the result and discussion of our previous experiment PALS work, the average $v_f$ will coalesce to larger holes.

There are imperfections of the model:

1. The molar volume doesn't linearly expand with temperature exactly for some simple liquids. First order approximation of the curve is utilized in this work.

2. The first order approximation is applied when calculate the volume and potential energy change.

3. It is questionable to use $V_w$ as hard core volume. Precise work is needed for this part.

4. The shape factor is not considered in this work, more detailed work is needed in the future study.

5. Lennard-Jones potential only estimation of forces between particles.

*Acknowledgements*

*Dr. Yang Yu acknowledges the financial support from the National Natural Science Foundation of China (Grant No.: 11247220) and the Science Research Startup Foundation from the Nanjing University of Information Science & Technology (Grant No: S8112078001). Supported by the Natural Science Foundation of Jiangsu Province (No. BK20131428 )*